\documentclass{elsart}

\usepackage{natbib}

 \usepackage{graphicx}
\usepackage{amssymb}

\def\ks{km s$^{-1}$}
\def\cm3{cm$^{-3}$}

\def\m{$^\prime$}
\def\s{$^{\prime\prime}$}

\begin{document}

\begin{frontmatter}



\title{The neutral gas in the environs of the Geminga gamma-ray pulsar}


\author[label1]{E. Giacani\corauthref{cor1}}
\ead{egiacani@iafe.uba.ar}
\author[label1]{E. M. Reynoso\corauthref{cor1}}
\author[label1]{G. Dubner\corauthref{cor1}}
\author[label2]{W. M. Goss}
\author[label3]{A. J. Green}
\author[label3]{S. Johnston}
\corauth[cor1]{Member of the Carrera del Investigador Cient\'\i fico, CONICET}

\address[label1]{Instituto de Astronom\'{\i}a y  F\'{\i}sica del Espacio 
(IAFE), CC 67, Suc. 28, 1428 Buenos Aires, Argentina}
\address[label2]{National Radio Astronomy Observatory, P.O.
Box 0, Socorro, New Mexico 87801, USA}
\address[label3]{School of Physics, University of Sydney, NSW 2006, Australia}
%\address[label4]{Australia Telescope National Facility, CSIRO, P.O. Box 76, 
%Epping, NSW 1710, Australia}
\begin{abstract}
We present a  high-resolution (24\s) study of the HI interstellar gas 
distribution around the radio-quiet neutron star Geminga. 
Based on Very Large Array (VLA)  and MPIfR
Effelsberg telescope data, we analyzed a 40\m$\times$40\m~ field around
Geminga. These observations have revealed the presence of a neutral
gas shell, 0.4 pc in radius, with an associated HI mass of 0.8 M$_{\odot}$, 
which surrounds Geminga at a radial velocity compatible with the kinematical 
distance of the neutron star. In addition, morphological agreement is 
observed between the internal face of the HI shell and the brightest 
structure of Geminga's tail observed in X-rays. 
We explore the possibility that this morphological agreement is the result
of a physical association. 
\end{abstract}

\begin{keyword}
neutron stars \sep neutron stars: Geminga \sep neutral hydrogen


\end{keyword}

\end{frontmatter}
% main text
\section{Introduction}
\label{}

Neutron stars (NSs) have generally been detected as radio pulsars.
In recent years, with the advent of high-energy instruments, several new
classes of NSs have been discovered as unresolved X-ray sources. These
include: anomalous X-ray pulsars, soft gamma ray repeaters, isolated
neutron stars (also called ``radio quiet neutron stars''), and compact
central objects when the NSs are located inside supernova remnants (SNR).
We have 
undertaken an investigation of the influence that these peculiar sources may 
have on the neutral gas in their environment. We have already studied three
NSs: 1E 1207-5209 in G296.5+10.0 \citep{Giacani00}, RX J0822--4300 in
Puppis A \citep{Reynoso03}, and 1E 161345-5055 in RCW 103 \citep{Reynoso04}.
In all three cases there is a minimum in the HI emission, coincident with 
the location of the X-ray source, and at a radial velocity corresponding 
to the NS's kinematical distance. Hence, an association between the sources and 
their corresponding HI feature is quite probable. However, the physical 
mechanism that might give rise to such a phenomenon in the neutral HI gas 
remains unresolved.

We present here results based on HI observations obtained for the isolated 
Geminga pulsar (PSR J0633+1746). This source is known to pulsate from 
$\gamma$-rays to the optical band. In the radio domain, three groups 
independently reported the detection of weak pulsed radio emission
at 102 MHz \citep{Malofeev97, Kuzmin97, Shitov98}. However, the detection of a 
radio continuum counterpart at other frequencies (74 and 330 MHz) have so 
far been negative (\citealt{Kassim99} and references therein). Other groups 
have also failed to detect radio pulsations (Hankins, private communication).
Geminga has a spin-down age of 3.4$\times 10^{5}$ yr and a rotational
energy loss of $\dot E=3.2 \times 10^{34}$ erg s$^{-1}$ \citep{Bertsch92}. 
 A  distance of 160 pc and a transverse velocity of
130 \ks~  were determined from optical parallax measurements using the 
Hubble Space Telescope \citep{Caraveo96}.
Recently, \citet{Caraveo03} reported 
the XMM-Newton X-ray detection of two elongated X-ray tails originating close 
to Geminga, which run parallel to the direction of motion of the pulsar and 
extend about 2\m. The emission is produced by particles expelled
by Geminga's strong magnetic field. The tails are the bright edges of
the three-dimensional shockwave created by Geminga's motion. It has been
suggested that Geminga is travelling almost directly transverse to the line of
sight \citep{Caraveo03}. 

To constrain the physical properties of Geminga and its associated 
X-ray nebula, we have investigated their possible interactions with
the surrounding interstellar HI gas.

\section {Observations}

A field of 40$^\prime \times 40^\prime$ was observed in the HI ($\lambda$ 21
cm line) with the VLA\footnote{The VLA of the NRAO is a facility of the NSF, 
operated under cooperative agreement by Associated Universities, Inc.} between
2001 September 3 and 2001 November 20 in the C, CnD and D configurations 
for a total of more than 15 hours. The  observations used the correlator in 
single-IF mode with on-line Hanning smoothing. Spectral information was detected 
in 256 channels, with a velocity resolution of 
0.64 \ks, over the velocity range from --82 to +82 \ks~(all velocities
are relative to the LSR). Data reduction and calibration were carried out 
using the AIPS software package, following standard procedures. The $u-v$
visibilities from all the datasets were combined and the complete 
interferometric data cube has a 1$\sigma$ rms noise of $\sim$2 mJy beam$^{-1}$.
 
To recover short spatial frequencies, the VLA data were combined in the 
$u-v$ plane with single dish observations from the 100 m MPIfR radio telescope
at Effelsberg. A 201 channel correlator was used with a total effective
bandwidth of 1.2 MHz, giving a velocity resolution of 1.3 \ks~ and an rms noise 
per channel of about 40 mK. These data are in units of brightness temperature, 
calibrated against the IAU standard position S7 ($l = 132^{\circ}, 
b = -1^{\circ}$; \citet{Kalberla80}). To combine the data sets, a factor of 1052 
was used to convert the VLA flux densities from Jy beam$^{-1}$ to brightness 
temperature in Kelvin. The angular resolution of the combined image is 24\s~
and the 1$\sigma$ rms noise in line-free channels is about 0.4 K. The final 
cube covers the velocity range --68 to +68 \ks~ with a velocity resolution 
of 1.3 \ks.

\section {Results and Discussion}

We have examined the HI data cube looking for perturbations in the interstellar
 medium (ISM) that could be caused by the presence of the NS. HI
emission features appear only in three velocity intervals.  Most of the 
emission is near 15 \ks, and is associated with the local spiral arm. 
The second  emission peak is near 22 \ks, which can be interpreted as emission from
gas related to the Perseus spiral arm \citep{Vogt76, Caswell87}. 
The only feature that could be related to Geminga appears in the third velocity 
range, between 1.3 and 4.5 \ks. 

\begin{figure}
      \centering
\vspace{1cm}
      \includegraphics [width=8cm]{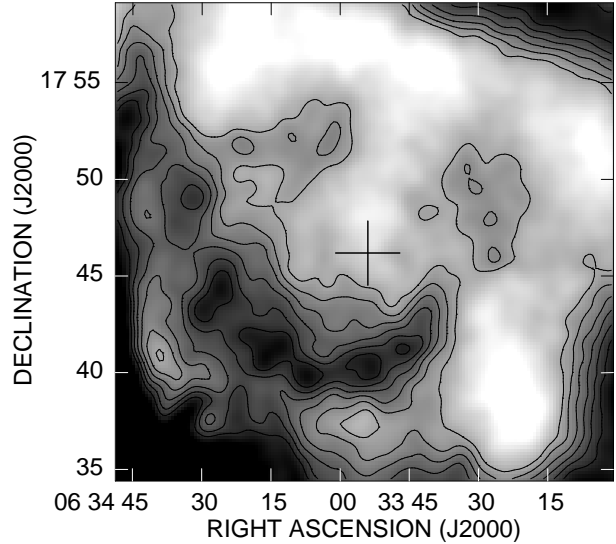}
      \caption{Gray-scale and contour image of the HI emission towards
Geminga averaged over the velocity interval 1.3 -- 4.5 \ks. The gray-scale 
ranges from 30 to 50 K, while the contour levels are from 37 to 47 K in steps 
of 2 K. The angular resolution has been smoothed to 50\s~ to facilitate the 
identification of intense features. The cross indicates the position of
Geminga.}
\end{figure}

At this third velocity range,  Geminga is surrounded by an open 
HI shell 
(Figure 1). This incomplete 
ring, with an average angular radius of 9\m~ (0.4 pc at 160 pc), 
opens to the NW. The neutron star appears projected onto an HI void,  which 
has a minimum at RA = 06$^{\rm h}~33^{\rm m}~ 53.0^{\rm s}$, Dec = 
17$^{\circ}$ 47\m 47\s
 (J2000), but is offset by about 2\m~ (0.09 pc) from the hole center.  The
close proximity of Geminga to both the Earth and the direction of the Galactic 
anticenter precludes a reliable estimate of the kinematical distance using the 
Galactic circular rotation model. However, the presence of this HI structure at
such low velocities is consistent with the distance for Geminga, estimated by 
independent methods, and thus with a possible association. Moreover, the 
fact that Geminga is evolving inside a bubble is in good agreement
with the very low ambient density values ($\sim$ 0.06 to 0.15 cm$^{-3}$ 
derived by Caraveo et al. 2003) to
confine the X-ray bow-shock structure. 

We have estimated the HI mass content of the shell by integrating the column 
density distribution within the velocity interval (1.3, 4.5) \ks, assuming
that the gas is optically thin. A total of 0.8 M$_{\odot}$ is obtained. 
It is difficult to explain the  origin  of the HI feature in connection with 
Geminga. Taking into account the proper motion of the pulsar
and its age (t$\simeq 340000$ yrs), Geminga was created in a supernova 
explosion 45 pc away from its current position. Therefore  any possible
association between  the HI shell and Geminga's parent supernova can be 
ruled out.

In Figure 2a we display a comparison of the HI gas integrated between 1.3 and 
4.5 \ks~ with the X-ray emission in the range 0.3 to 5 KeV from 
\citet{Caraveo03}. There is excellent morphological agreement between the 
southern tail of the associated X-ray nebula and the internal border of 
the HI shell. This tail is the brighter of the two extensions in the X-ray 
source. It is possible that the higher HI density in this region has 
contributed to an enhancement of this X-ray feature.

\begin{figure}
      \centering
\vspace{1cm}
      \includegraphics [width=14cm]{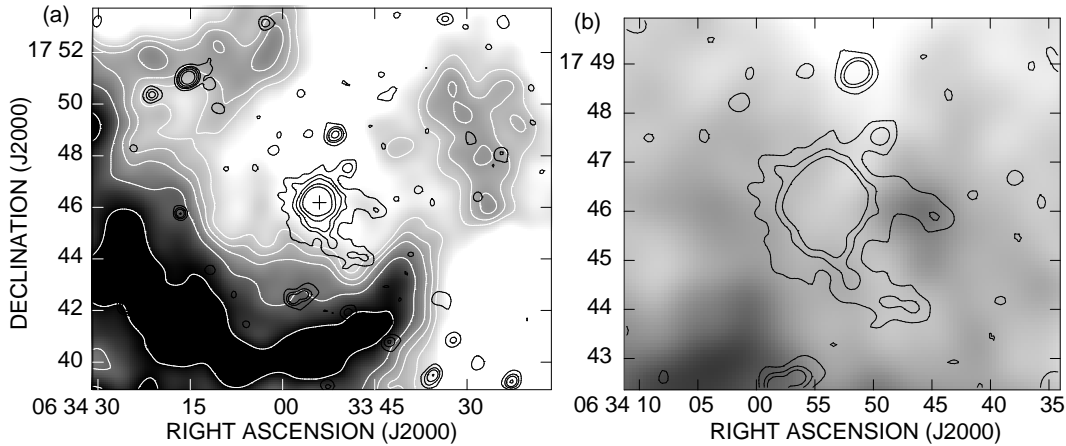}
      \caption{(a) XMM-Newton X-ray data in the range 0.3 to 5 KeV (black
contours) from Caraveo et al. 2003, superimposed onto the integrated HI
emission between 1.3 and 4.5 \ks~ (gray-scale and white contours).
The cross indicates the position of Geminga.
(b) The HI column density, integrated between 0 and 2 \ks, is shown
in gray-scale and ranges from (1.0 -- 1.6) $\times 10^{20}$ cm$^{-2}$. To show
the location of the X-ray source, a few contours from the image of 
\citet{Caraveo03} are included.}
\end{figure}

The fact that the only HI minimum observed along the whole  
 data cube 
agrees spatially and in velocity with  Geminga's location, together with the 
excellent morphological agreement with the X-ray nebula, reinforces the 
possibility of an association. We have conducted a literature search for 
early-type stars, open clusters, OB associations and HII regions that could  
have created HI shell-type features in the region, but none were found.
Also, inspection of the IRAS (HIRES) images shows no infrared counterpart for 
this shell.

Figure 2b displays the neutral hydrogen column density  distribution,
N$_{\rm H}$,  obtained from the integration of the foreground HI
($v \sim$ 0 to 2 \ks).  The absorbing column density slightly varies from 
1.3 $\times 10^{20}$ cm$^{-2}$ in the direction of Geminga, to 1.5 $\times 
10^{20}$ cm$^{-2}$ towards the northern X-ray tail. It is 
significant that the HI column density does not increase between the tails,
indicating that the observed morphology of the X-ray source is real and not
a consequence of variable X-ray absorption due to the intervening HI column.

Based on the line-free channels of our 1.4 GHz cube,  we have produced  
an image of the radio continuum at this frequency in the direction of Geminga 
with 24\s~ angular resolution. We have not detected any radio continuum
source, either point-like or extended, down to a noise level of 0.14 
mJy beam$^{-1}$. We can speculate that if a radio pulsar wind nebula (PWN)
existed, at least the same size as the X-ray nebula, then the
maximum flux density at 1.4 GHz, consistent with our non-detection (at the
3$\sigma$ noise level) would be 54 mJy. For a typical PWN
spectral index of $\alpha = -0.3$ (S $\propto \nu^{\alpha}$), the broadband
radio luminosity L$_{\rm r}$ between $ 10^7$ and $10^{11}$ Hz would be
6$\times 10^{28}$ erg s$^{-1}$, corresponding to an efficiency $\epsilon  
\equiv $L$_{\rm r}/\dot E \sim 10^{-6}$. This value is significantly lower than 
those found for other radio PWN but
comparable to that of PSR B1706-44 \citep{Giacani01}, which is a young
isolated pulsar with a pulsed gamma-ray source. Since Geminga is an old pulsar,
the production efficiency of synchrotron radiation would be reduced. 
 We have also processed archival VLA data 
from B-array observations. The image, with a resolution of 4\s~ and an rms 
noise of 31 $\mu$Jy beam$^{-1}$, shows no counterpart. 

Summing up , the present work is the first high resolution, high sensitivity 
survey of the neutral interstellar gas around Geminga. These observations
have revealed that the pulsar lies inside  an interstellar bubble, probably
preexistent.    
The good morphological agreement between the brightest structure 
of Geminga's X-ray tail and the wall of the bubble suggests the existence
of interaction between them. 
A  theoretical investigation 
is necessary to understand the mechanism through which the HI/X-ray interaction
have enhanced the X-ray tail.

\medskip

The authors would like to thank Peter Kalberla for carrying out the
observations with the 100 m telescope of MPIfR (Max-Planck- Institut
f\"ur Radioastronomie) at Effelsberg and for processing those data.  
This research was funded through CONICET (Argentina) grant PIP2136/00 and
grant UBACYT (Argentina) A055.

% \appendix

% \section{}
% \label{}

% Bibliographic references with the natbib package:
% Parenthetical: \citep{Bai92} produces (Bailyn 1992).
% Textual: \citet{Bai95} produces Bailyn et al. (1995).
% An affix and part of a reference:
%   \citep[e.g.][Ch. 2]{Bar76}
%   produces (e.g. Barnes et al. 1976, Ch. 2).

\end{document}